\documentstyle[prd,aps,twocolumn,floats]{revtex}
\renewcommand{\epsilon}{\varepsilon}
\begin{document}
\twocolumn[\hsize\textwidth\columnwidth\hsize\csname
@twocolumnfalse\endcsname
\preprint{}
\title{Inflation in a self-interacting gas universe}
\author{Winfried Zimdahl\footnote{Electronic address: winfried.zimdahl@uni-konstanz.de}}
\address{Fakult\"at f\"ur Physik, Universit\"at Konstanz, PF 5560 M678
D-78457 Konstanz, Germany}
\author{Alexander B. Balakin
\footnote{Electronic address: Alexander.Balakin@ksu.ru}} 
\address{Fakult\"at f\"ur Physik, Universit\"at Konstanz, PF 5560 M678
D-78457 Konstanz, Germany\\
and 
Department of General Relativity and Gravitation,  
Kazan State University, 420008 Kazan, Russia\thanks{Present address}}
\date{\today}
\maketitle
\begin{abstract}

We show that a de Sitter spacetime is a solution of Einstein's field equations with the energy momentum tensor of a  
self-interacting, classical Maxwell-Boltzmann gas in collisional equilibrium.  The self-interaction is described by a four-force which is quadratic in the (spatially projected) particle four-momenta. 
This force does not preserve the particle number and gives rise to an exponential increase in the comoving entropy of the universe while the 
temperature of the latter remains constant. 
These properties of a gas universe are related to the existence of a 
``projector-conformal'' timelike Killing vector representing a symmetry  
which is ``in between'' the symmetries characterized by a Killing vector and those characterized by a conformal Killing vector. 
\end{abstract}
\vspace{1.5cm}
]

\section{Introduction}
Standard inflationary cosmology relies on the dynamics of a scalar field with a suitably designed potential term allowing for an effectively negative pressure of the cosmic substratum \cite{KoTu,Boe}. 
Alternatively, there have been numerous attempts to describe early phases of accelerated expansion, either power-law or exponential, as nonequilibrium (imperfect) fluid phenomena (see \cite{Roy,WZ} and references therein). 
Studies along this line make use of the fact that fluid viscosities are dynamically equivalent to effective negative pressures. 
Such kind of pressures occur, e.g., due to internal interactions inside relativistic gases \cite{WI,UI,ZMNRAS}. 
However, the dissipative fluid approaches face the following general problem. 
Their reliability is restricted to small deviations from perfect fluid behaviour, i.e., close to (local) equilibrium, while inflation in this context is necessarily  a far-from-equilibrium phenomenon. 
Although the imperfect fluid dynamics was shown to admit inflationary solutions  
(see \cite{Roy,WZ} and references therein) their physical significance is unclear \cite{RM}. 
The question whether or not bulk viscosity may drive inflation \cite{Pach}
remains open. 

A different line of fluid dynamical early universe investigations implying effective negative pressures as well is connected with cosmological particle production 
\cite{Prig,Barr,Calv,LiGer,ZP1,ZPJ,ZP2,GaLeDe,Lima,TZP,ZTP,WZMNRAS,ZPM}. 
Here, the negative bulk pressure is not due to deviations from (collisional) equilibrium but is a consequence of the phase space enlargement of the fluid particle system. 
Negative pressure terms of this kind in the local energy-momentum conservation relations may be regarded as equivalent to corresponding not necessarily small source terms in the energy-momentum balance of a perfect fluid. In other words, the effective nonequilibrium description in terms of imperfect fluid quantities mimics a perfect fluid with continuously increasing particle number, supposedly of quantum origin.  
While the magnitude of conventional viscous pressures which are due to particle number preserving interactions within the fluid is severely restricted by close-to-equilibrium conditions, the magnitude of effective, cosmological particle production describing bulk pressures is not. 

Previous investigations have shown that such kind of (generalized) equilibrium particle production (creation of perfect fluid particles with minimal entropy production) may substantially modify the standard cosmological dynamics, including the possibility of ``reheating'' phenomena and power-law inflation \cite{ZTP,ZPM,Zsub,ZiBa}. 
Particularly interesting connections were established between spacetime symmetries, described by a conformal timelike Killing-vector and the production rate for particles at equilibrium. 
Depending on the equations of state the symmetry requirements turned out to fix the creation rate for fluid particles at equilibrium. 
Moreover, for a gas universe the production of particles could be traced back to specific self-interacting forces on the (classical) microscopic constituents of the cosmic medium \cite{ZiBa}. 
Essential features of the cosmological dynamics may be discussed in terms of microscopic particle motion in a (classical) force field. 

In section III of the present paper we identify those forces on the particles of a simple relativistic gas governed by an equilibrium distribution function which, on the phenomenological level, give rise to an effective bulk pressure $\pi = - \left(\rho + p \right)$, where $\rho $ is the energy density of the cosmic medium and $p$ is the corresponding equilibrium pressure.  
A negative pressure $\pi $ of this magnitude is equivalent to a particle production rate that coincides with the expansion rate of the universe 
(see Eq. (\ref{17}) below). 
We show that a gas configuration such as this is intimately connected with a symmetry of the spacetime characterized by a ``projector-conformal timelike Killing-vector'' 
(PCTKV) $\zeta ^{a}$ which we generally define by  
\begin{equation}
\pounds _{_{\zeta _{a}}} g _{ik} \equiv  \zeta _{i;k} + \zeta _{k;i} 
= 2\phi P_{ik}\ ,
\label{1}
\end{equation}
where $\pounds _{_{\zeta _{a}}} g _{ik}$ is the Lie-derivative of the metric tensor $g _{ik}$ with respect to $\zeta _{a}$ and 
$\phi = \phi \left(x \right) $ is a spacetime function.
The quantity $P _{ik}$ 
is the projector 
\begin{equation}
P _{ik} = g _{ik} - \frac{\zeta _{i}\zeta _{k}}
{\zeta _{a}\zeta ^{a}}\ ,
\mbox{\ \ \ }
P _{ik}\zeta ^{k} = 0
\label{1a}
\end{equation} 
on surfaces orthogonal to the timelike vector $\zeta ^{a}$. 
The special case of interest here corresponds to $\zeta _{a} = u _{a}/T $,  
where $u ^{a}$ is the macroscopic four-velocity of the cosmic medium and 
$T$ is its equilibrium temperature. 
The projector $P _{ik}$ then reduces to  
$h _{ik} = g _{ik} + u _{i}u _{k}$, the projection tensor orthogonal 
to $u ^{i}$. 
The first part of the paper (section II) therefore intends to clarify the general implications of the PCTKV property of $u _{a}/T$ for the 
relativistic fluid dynamics. 
Section IV considers the case that a medium such as characterized in section III dominates the dynamics of a homogeneous and isotropic universe, while section V 
summarizes our main findings.  
Units have been chosen so that $c = k_{B} = \hbar = 1$. 

\section{Fluid dynamics with a ``projector-conformal'' 
timelike Killing-vector (PCTKV)}
In fluid spacetimes the Lie derivative $\pounds _{_{\frac{u _{a}}{T}}}g _{ab}$ of the metric tensor $g _{ab}$ with respect to $u _{a}/T$,  
where $u ^{a}u _{a} = -1$, plays a well-known role to characterize symmetries of the metric. 
Especially interesting are cases in which $u _{a}/T$ is either a Killing vector or a conformal Killing vector. 
Like any symmetric tensor, 
the Lie derivative of the metric may generally be split into contributions parallel and perpendicular to the four-velocity:
\begin{equation}
\pounds _{_{\frac{u _{a}}{T}}} g _{ik} \equiv   \left(\frac{u _{i}}{T} \right)_{;k} 
+ \left(\frac{u _{k}}{T} \right)_{;i} = C _{ik}  
\label{1b}
\end{equation}
with 
\begin{equation}
C _{ik} = 2 A u _{i}u _{k} + B _{i}u _{k} + B _{k}u _{i} 
+ 2 \phi h _{ik} + b _{ik}
\label{2}
\end{equation}
and $h _{ik}u ^{k} = B _{i}u ^{i} = b _{ik}u^{k} = b ^{i}_{i} = 0$. 
Combining relations (\ref{1b}) and (\ref{2}) we have 
\begin{equation}
A = \frac{1}{2}C _{ik}u ^{i}u ^{k} = \frac{\dot{T}}{T ^{2}}\ ,
\label{2a}
\end{equation}
\begin{equation}
B _{m} = - h ^{i}_{m}u ^{k}C _{ik} 
= - \frac{1}{T}\left[\dot{u}_{m} + \frac{\nabla  _{m}T}{T} \right]\ ,
\label{2b}
\end{equation}
\begin{equation}
\phi = \frac{1}{6}h ^{ik}C _{ik} =  \frac{1}{3}
\frac{\Theta }{T}\ ,
\label{2c}
\end{equation}
and 
\begin{eqnarray}
b _{ab} &=& \left[h _{a}^{i}h _{b}^{k} - \frac{1}{3}h _{ab}h ^{ik} \right]
C _{ik}\nonumber\\
&=& \frac{1}{T}\left[\nabla  _{b}u _{a} + \nabla  _{a}u _{b} 
- \frac{2}{3}\Theta h _{ab}\right] 
\equiv  \frac{2 \sigma _{ab}}{T}\ ,
\label{2d}
\end{eqnarray}
where $\Theta \equiv  u ^{i}_{;i}$ is the fluid expansion, 
$\dot{T} \equiv  T _{,a}u ^{a}$, $\nabla  _{a}T \equiv  h _{a}^{b}T _{,b}$, 
$\dot{u}_{a}$ is the four-acceleration and 
$\sigma _{ab}$ is the shear. 

According to different choices for the quantities $A$, $B _{m}$, $\phi $, and 
$b _{ab}$ we introduce the following classification for $u _{a}/T$: 
%\begin{itemize}

%\item
(i) $u _{a}/T$ is a Killing-vector for $A = B _{m} = \phi = b _{ab} = 0$, corresponding to 
\begin{equation}
\left(\frac{u _{i}}{T} \right)_{;k} 
+ \left(\frac{u _{k}}{T} \right)_{;i} = 0  \ ,
\label{2e}
\end{equation}
or   
\begin{equation}
\Theta = \dot{T} = 0 \ ,
\mbox{\ \ \ }
\frac{\nabla  _{a}T}{T} + \dot{u}_{a} = 0 \ ,
\mbox{\ \ \ }
\sigma _{ab} = 0 \ .
\label{2f}
\end{equation}

%\item
(ii) $u _{a}/T$ is a conformal Killing-vector for $A = - \phi$ and 
$B _{m} = b _{ab} = 0$, i.e.  
\begin{equation}
\left(\frac{u _{i}}{T} \right)_{;k} 
+ \left(\frac{u _{k}}{T} \right)_{;i} = 2 \phi g _{ik}  \ ,
\label{2g}
\end{equation}
equivalent to relations (\ref{2c}) and 
\begin{equation}
\frac{\dot{T}}{T} = - \frac{\Theta }{3}\ ,
\mbox{\ \ \ }
\frac{\nabla  _{a}T}{T} + \dot{u}_{a} = 0 \ ,
\mbox{\ \ \ }
\sigma _{ab} = 0 \ .
\label{2h}
\end{equation}

%\item
(iii) $u _{a}/T$ is a ``projector-conformal'' timelike Killing-vector (PCTKV) 
for  $A = B _{m} = b _{ab} = 0$, i.e., 
\begin{equation}
\left(\frac{u _{i}}{T} \right)_{;k} 
+ \left(\frac{u _{k}}{T} \right)_{;i} = 2 \phi h _{ik}  \ ,
\label{5}
\end{equation}
or relations (\ref{2c}) and 
\begin{equation}
\frac{\dot{T}}{T} = 0\ ,
\mbox{\ \ \ }
\frac{\nabla  _{a}T}{T} + \dot{u}_{a} = 0 \ ,
\mbox{\ \ \ }
\sigma _{ab} = 0 \ .
\label{6}
\end{equation}

%\end{itemize}

In this paper we are interested in the third of these cases. 
In the following section we will give a microscopic realization of the PCTKV  behavior on the level of kinetic theory and identify a spacetime which admits a PCTKV. 

In the present section we show that the property (\ref{2c}) with a nonvanishing $\Theta $ together with Eqs. (\ref{6}) may be realized in case the fluid particle number is not preserved. 

Introducing a length scale $a$ according to $\Theta = 3 \dot{a}/a$, the number $N$ of particles in a comoving volume $a ^{3}$ is $N = na ^{3}$, 
where $n$ is the particle number density. 
The corresponding particle number flow vector is $N ^{a} = n u ^{a}$. 
Denoting the phenomenological particle production rate by $\Gamma $, the particle number balance may be written as  
\begin{equation}
N ^{a}_{;a} = \dot{n} + \Theta n = n\frac{\dot{N}}{N} = n \Gamma \ .
\label{7}
\end{equation}

The energy-momentum tensor of a general fluid is given by 
\begin{equation}
T ^{ik} = T ^{ik}_{\left(0 \right)} + \pi  h ^{ik} 
+ \pi ^{ik} + q ^{i}u ^{k} + q ^{k}u ^{i}
\label{8}
\end{equation}
with 
$T ^{ik}_{\left(0 \right)} = \rho  u ^{i}u ^{k} + p h ^{ik}$
and 
$\pi ^{ik}u _{k} = q ^{i}u _{i} = 
\pi ^{i}_{i} =  h ^{ik}u _{i} = 0$. 
Local energy-momentum conservation $T ^{ik}_{\ ;k} = 0$ implies 
\begin{equation}
\dot{\rho } = - \Theta \left(\rho + p + \pi  \right) 
- \nabla _{a}q ^{a} - 2 \dot{u}_{a}q ^{a} 
- \sigma _{ab}\pi ^{ab}\ .
\label{9}
\end{equation}
With the help of the Gibbs equation (see, e.g., \cite{Groot})
\begin{equation}
T \mbox{d}s = \mbox{d} \frac{\rho }{n} + p \mbox{d}\frac{1}{n}\ ,
\label{10}
\end{equation}
where $s$ is the entropy per particle, we obtain 
\begin{equation}
n T \dot{s} = \dot{\rho } - \left(\rho + p \right)
\frac{\dot{n}}{n}\ .
\label{11}
\end{equation}
Taking into account the standard definition 
$S ^{i} = n s u ^{i} + q ^{i}/T$ 
of the entropy flow vector $S ^{i}$  
and using Eqs. (\ref{7}), (\ref{9}), and (\ref{11}) yields 
\begin{eqnarray}
S ^{i}_{;i} - n s \Gamma  &=& 
- \frac{\rho + p}{T}\Gamma \nonumber\\
&&- \frac{1}{2}\left(T ^{ik} - T ^{ik}_{\left(0 \right)} \right)
\left[\left(\frac{u _{i}}{T}\right)_{;k} 
+ \left(\frac{u _{k}}{T}\right)_{;i}\right]   \ .
\label{12}
\end{eqnarray}
Under the PCTKV conditions (\ref{5}), Eq. (\ref{12}) reduces to 
\begin{equation}
S ^{i}_{;i} - ns \Gamma 
= - \frac{\rho + p}{T}\Gamma - \frac{\Theta \pi }{T} \ .
\label{13}
\end{equation}
For $q ^{a} = \pi ^{ab} = 0$ the right-hand side of this equation coincides with $n \dot{s}$. 
The especially interesting case 
$\Theta \pi = - \left(\rho + p \right)\Gamma $ which makes the right-hand side of Eq. (\ref{13}) vanish implies $S ^{i}_{;i} = ns \Gamma $ 
(and $\dot{s} = 0$ for $q ^{a} = \pi ^{ab} = 0$): There is entropy production only due to the enlargement of the phase space of the system but not to conventional dissipative processes within the fluid, i.e., the entropy production is minimal.  
States in which the right-hand side of Eq. (\ref{13}) vanishes 
(and $\Gamma \geq 0$ holds) are states of ``generalized equilibrium''
\cite{Zsub}. 
This kind of equilibrium was originally introduced on the basis of the 
conformal Killing vector property (\ref{2g}) of $u _{i}/T$.  
Here we enlarge the generalized equilibrium concept to include the PCTKV case 
(\ref{5}) as well. 
{\it We consider a fluid to be in ``generalized equilibrium'' if 
(i) $S ^{i}_{;i} - ns \Gamma = 0$ with $\Gamma \geq 0$ holds and 
(ii) $u _{i}/T$ satisfies either the conditions (\ref{2g}), equivalent to Eqs. (\ref{2c}) and (\ref{2h}), or the conditions (\ref{5}), equivalent to Eqs. (\ref{2c}) and (\ref{6})}. 

For massive particles in a homogeneous and isotropic universe 
the first case, minimal entropy production under the conditions (\ref{2g}), dealt with in \cite{Zsub,ZiBa}, was shown to imply power-law inflation. 
In this paper we are interested in the second case, i.e., minimal entropy production according to $S ^{i}_{;i} - ns \Gamma = 0$ together with the PCTKV property (\ref{5}). We will demonstrate that this type of generalized equilibrium under the conditions of homogeneity and isotropy 
requires a de Sitter universe for arbitrary fluid equations of state. 
 
It is a characteristic feature of generalized equilibrium of both types 
that the particle production rate is not an arbitrary parameter but determined by consistency requirements. 
Given equations of state in the general form 
\begin{equation}
p = p \left(n,T \right)\ ,\ \ \ \ \ \ \ \ \ 
\rho = \rho \left(n,T \right)\ ,
\label{14}
\end{equation}
differentiation of the latter relation and using 
the balances (\ref{7}) and  
(\ref{9}) yields 
\begin{equation}
\frac{\dot{T}}{T} = - \left(\Theta - \Gamma  \right) 
\frac{\partial p}{\partial \rho } 
+ \frac{n \dot{s}}{\partial \rho / \partial T}\ ,
\label{15}
\end{equation}
where the abbreviations
\[
\frac{\partial{p}}{\partial{\rho }} \equiv  
\frac{\left(\partial p/ \partial T \right)_{n}}
{\left(\partial \rho / \partial T \right)_{n}} \ ,
\ \ \ \ \ \ \ \ 
\frac{\partial{\rho }}{\partial{T}} \equiv  
\left(\frac{\partial \rho }{\partial T} \right)_{n}
\]
have been used. 
Restricting ourselves to the case $\dot{s} = 0$,  
the temperature laws in Eqs. (\ref{6}) 
and (\ref{15}) are only consistent for $\Gamma = \Theta$. 
Via the relation 
\begin{equation}
\Theta \pi = - \left(\rho + p \right)\Gamma  
\label{16}
\end{equation}
which makes the right-hand side of Eq. (\ref{13}) vanish,  
the viscous pressure $\pi $ is fixed as well:
\begin{equation}
\Gamma = \Theta \quad \Rightarrow \quad \pi = - \left(\rho + p \right)\ .
\label{17}
\end{equation}
From Eq. (\ref{7}) follows  that $\dot{n}$ vanishes. 
Because of the equations of state (\ref{14}), constant values of $n$ 
and $T$ along the fluid flow lines imply constant pressure and constant energy density as well:
\begin{equation}
\frac{\dot{n}}{n} = \frac{\dot{T}}{T} = 0
\quad \Rightarrow \quad \dot{p} = \dot{\rho } = 0 \ .
\label{17a}
\end{equation}

We recall \cite{Zsub} that generalized equilibrium under the conformal Killing-vector conditions (\ref{2g}) is characterized by Eq. (\ref{16}) with
\[
\Gamma = \left(1 - \frac{1}{3}\frac{\partial{p}}{\partial{\rho }} \right)
\Theta \ ,
\mbox{\ \ \ }
\pi = - \left(\rho + p \right)
\left(1 - \frac{1}{3}\frac{\partial{p}}{\partial{\rho }}\right)
\]
instead of Eq. (\ref{17}). 

Statements on the spatial dependences of the thermodynamic quantities may be obtained from the momentum balance 
\begin{equation}
\left(\rho + p + \pi  \right)\dot{u} _{m} 
+ \nabla  _{m}\left(p + \pi  \right) = 0 \ ,
\label{18}
\end{equation}
where we restricted ourselves again to $q ^{a} = \pi ^{ab} = 0$.   
Because of relation (\ref{17}) we find 
\begin{equation}
\nabla  _{a}p = - \nabla  _{a}\pi 
\quad \Rightarrow \quad \nabla  _{a}\rho = 0 \ .
\label{19}
\end{equation}
The energy density is also spatially constant, the pressure not necessarily. 
With the help of the second equation of state (\ref{14}) the condition 
(\ref{19}) provides us with a relation between $\nabla  _{a}T$ and 
$\nabla  _{a}n$:
\begin{equation}
\frac{\nabla  _{a}T}{T} = - \frac{n}{T}\frac{ \partial \rho / \partial n}
{ \partial \rho / \partial T}\frac{\nabla  _{a}n}{n}\ ,
\label{20}
\end{equation}
where 
\[
\frac{\partial{\rho }}{\partial{n}} = \frac{\rho + p}{n} 
- \frac{T}{n}\frac{\partial{p}}{\partial{T}}\ ,
\]
which is a consequence of the fact that the entropy is a state function. 

Restricting ourselves to a classical gas with $p = nT$ and using 
the first relation (\ref{19}) and Eq. (\ref{20}) as well as the second relation (\ref{6}) we obtain
\begin{eqnarray}
\frac{\nabla  _{m}\pi }{nT} &=& \left[\frac{T}{n}
\frac{ \partial \rho / \partial T}{ \partial \rho / \partial n} - 1\right]
\frac{\nabla  _{m}T}{T} \nonumber\\
&=& - \left[\frac{T}{n}
\frac{ \partial \rho / \partial T}{ \partial \rho / \partial n} - 1\right]
\dot{u}_{m}\ .
\label{21}
\end{eqnarray}
From the Gibbs-Duhem relation 
\begin{equation}
\mbox{d} p = \left(\rho + p \right)\frac{\mbox{d} T}{T} 
+ n T \mbox{d} \left(\frac{\mu }{T} \right)
\label{22}
\end{equation}
follows under such circumstances that 
\begin{equation}
\nabla  _{m}\left(\frac{\mu }{T} \right) = 
- \left[\frac{\rho }{p} + \frac{T}{n}\frac{ \partial \rho / \partial T}
{ \partial \rho / \partial n} \right]
\frac{\nabla  _{m}T}{T}\ .
\label{23}
\end{equation}
 
The Gibbs equation (\ref{10}) together with the second relation (\ref{19}) then provides us with  
\begin{eqnarray}
\nabla  _{a}s &=& - \left(\frac{\rho }{p} + 1 \right)
\frac{\nabla  _{a}n}{n} \nonumber\\
&=& - \left(\frac{\rho }{p} + 1 \right)
\frac{T}{n}\frac{ \partial \rho / \partial T}
{n\partial \rho / \partial n}
\ \dot{u}_{a}\ .
\label{24}
\end{eqnarray}
The spatial gradient of the entropy per particle does not vanish in general.

We finish this section by considering the general relations (\ref{19}) - (\ref{24}) for the limiting cases of 
pure radiation (ultrarelativistic matter) and nonrelativistic matter. 

(i) $p = nT$, $\rho = 3 nT$ (ultrarelativistic matter):\\
The relation $\nabla  _{m}\rho = 0$ implies 
$\nabla  _{m}T/T = - \nabla  _{m}n/n$, equivalent to 
$\nabla  _{m}p = - \nabla  _{m}\pi = 0$. 
From Eq. (\ref{23}) one obtains
\begin{equation}
\nabla  _{m}\left(\frac{\mu }{T} \right) 
= - 4 \frac{\nabla  _{m}T}{T} = 4 \dot{u}_{m}\ ,
\ \ \ \ \ \ \ 
\left(p = \frac{\rho }{3}  \right)\ .
\label{25}
\end{equation}
Although the spatial pressure gradient vanishes, the corresponding gradients of the temperature, the number density and the entropy per particle 
are different from zero unless the fluid motion is geodesic. 

(ii) $p = nT$, $\rho = nm + \frac{3}{2}nT$, $m \gg T$ (nonrelativistic matter):\\ 
In this case the second relation (\ref{19}) reduces to 
\begin{equation}
\frac{\nabla  _{m}T}{T} \approx - \frac{2}{3}\frac{m}{T}
\frac{\nabla  _{m}n}{n}\ ,
\ \ \ \ \ \ \  
\left(m \gg T \right)\ ,
\label{26}
\end{equation}
and we find 
\begin{equation}
\nabla  _{m}p
 = - \nabla  _{m}\pi \approx - p\  \dot{u}_{m}\ ,
\ \ \ \ \ \ \ 
\left(m \gg T \right)\ .
\label{27}
\end{equation}
Nonrelativistic matter ($\rho \gg p$)  
in the present case combines a homogeneous energy density with a generally inhomogeneous equilibrium pressure. Only the total effective pressure is homogeneous as well. 

For the spatial gradient of $\mu /T$ we get 
\begin{eqnarray}
\nabla  _{m}\left(\frac{\mu }{T} \right) &\approx& 
- \frac{m}{T}\frac{\nabla  _{m}T}{T} 
= \frac{m}{T}\ \dot{u}_{m} \nonumber\\
&\quad \Rightarrow \quad &
\nabla  _{m}\left(\frac{\mu - m}{T} \right) \approx 0 \ ,
\ \ \ \ 
\left(m \gg T \right)\ .
\label{28}
\end{eqnarray}
It is well known that $\mu - m$ is the 
nonrelativistic chemical potential (see, e.g., \cite{Ehl}). 

In the following section we show how a gas characterized by the properties (\ref{16}) and (\ref{17}) may be realized with the help of relativistic kinetic theory.

\section{Kinetic theory for a gas in a force field}

\subsection{General relations}
The conventional kinetic theory of a simple relativistic gas relies on the concept of pointlike particles which may interact through elastic, binary collisions. 
Inbetween the collisions which are assumed to establish an 
(approximate) local or global equilibrium of the system the particles move on geodesics of either a given spacetime or a spacetime selfconsistently determined by the gas particles themselves. 
Sophisticated solution techniques for the corresponding Boltzmann equation 
have been developed and applied to numerous physically relevant situations \cite{Groot,Ehl,IS,Stew,MaWo}. 
The usual procedure here is first to characterize equilibrium states, i.e., states with vanishing entropy production and to relate the parameters of the corresponding distribution function to macroscopic (perfect) fluid quantities. 
Nonequilibrium situations are then, in a second step, taken into account as deviations from equilibrium. 
Obviously, geodesic particle motion is a highly idealized case. 
In reality, particle worldlines are supposed to deviate from geodesics 
since the particles will be subject to additional interactions in general. 
We assume here that these interactions may be modelled as effective forces on the particles. 
The kinetic theory for particles under the influence of various forces was considered, e.g.,  in \cite{Groot,Ehl,Hakim,BaGo}. 
Following the lines of \cite{ZiBa} 
we will focus here on equilibrium states of the gas in a 
force field. 
Our main objective will be the characterization of equilibrium configurations of a gas under the influence of a self-interacting force which is quadratic in the particle four-momenta.  
By ``self-interacting'' we mean that the force, except its dependence on the microscopic particle momenta, also depends on macroscopic quantities, characterizing the gas system as a whole (see below). 
This force will neither preserve the particle number nor the energy momentum and it will give rise to entropy production. 
In particular, it will turn out that a quadratically self-interacting force realizes states of ``generalized equilibrium''  characterized below Eq. (\ref{13}), including the condition $\dot{s} = 0$ for ``adiabatic'' 
(or ``isentropic'') particle production. 

The one-particle distribution function 
$ f = f\left(x,p\right)$ 
for  relativistic gas particles under the influence of a four-force 
$F ^{i}=F ^{i}\left(x,p \right)$ obeys the Boltzmann equation
\begin{equation} 
p^{i}f,_{i} - \Gamma^{i}_{kl}p^{k}p^{l}
\frac{\partial f}{\partial
p^{i}} + m F ^{i}\frac{\partial{f}}{\partial{p ^{i}}}
 = C\left[f\right]   \mbox{ , } \label{29}
\end{equation}
where $f\left(x, p\right) p^{k}n_{k}\mbox{d}\Sigma dP$ 
is the number of
particles whose world lines intersect the hypersurface element 
$n_{k}d\Sigma$ around $x$, having four-momenta in the range 
$\left(p, p + \mbox{d}p\right)$. 
The quantity $\mbox{d}P = A(p)\delta \left(p^{i}p_{i} + m^{2}\right) 
\mbox{d}P_{4}$
is the volume element on the mass shell 
$p^{i}p_{i} = -  m^{2} $
in the momentum space. 
$A(p) = 2$, if $p^{i}$ is future directed and 
$A(p) = 0$ otherwise; 
$\mbox{d}P_{4} = \sqrt{-g}\mbox{d}p^{0}\mbox{d}p^{1}
\mbox{d}p^{2}\mbox{d}p^{3}$. \\
$C[f]$ is Boltzmann's collision term. 
Its specific structure discussed e.g. by 
Ehlers \cite{Ehl} will not 
be relevant for our considerations. 
Following Israel and Stewart \cite{IS} we shall only 
require that (i) $C$
be a local function of the distribution function, i.e., 
independent
of derivatives of $f$, (ii) $C$ be consistent with 
conservation of
four-momentum and number of particles, and (iii) $C$
yield a nonnegative expression for the entropy 
production and do 
not vanish unless $f$ has the form of a local equilibrium
distribution function (see (\ref{35}) below). 
Equation (\ref{29}) implies that the mass-$m$ particles inbetween the 
collisions move according to the equations of motion
\begin{equation}
\frac{\mbox{d} x ^{i}}{\mbox{d} \gamma  } = p ^{i}\ ,
\ \ \ \ \ \ 
\frac{\mbox{D} p ^{i}}{\mbox{d} \gamma  } = m F ^{i}\ ,
\label{30}
\end{equation}
where $\gamma  $ is a parameter along their worldline which for massive particles 
may be related to the proper time $\tau $ by $\gamma   = \tau /m$. 
Since the particle four-momenta are normalized according to 
$p ^{i}p _{i} = - m ^{2}$, the force $F ^{i}$ has to satisfy the relation 
$p _{i}F ^{i} = 0$. 

Both the collision integral $C$ and the force $F ^{i}$ describe interactions within the many-particle system. 
While $C$ conventionally accounts for elastic binary collisions, 
we intend $F ^{i}$ to model different kinds of interactions in a simple manner. 
Strictly speaking, $F ^{i}$ should be calculated from the microscopic particle dynamics and, consequently, depend on the entire set of particle coordinates and momenta characterizing the system of gas particles.  
We will introduce here the simplifying assumption that  $F ^{i}$  be an effective one-particle quantity which instead of depending on the coordinates and momenta of the remaining particles, 
is supposed to depend on macroscopic fluid quantities characterizing the system as a whole. 
At the moment we do not specify this force. It will be determined below by general equilibrium conditions. 
We expect the concept of a self-interacting force to be useful in circumventing some of the general problems inherent in attempts to formulate a relativistic statistics for interacting many-particle systems (see, e.g., \cite{IsKan} and references therein). 

The particle number flow 4-vector 
$N^{i}$ and the energy momentum tensor $\tilde{T}^{ik}$ are
defined in a standard way (see, e.g., \cite{Ehl}) as 
\begin{equation}
N^{i} = \int \mbox{d}Pp^{i}f\left(x,p\right) \mbox{ , } 
\ \ \ 
\tilde{T}^{ik} = \int \mbox{d}P p^{i}p^{k}f\left(x,p\right) \mbox{ .} 
\label{31}
\end{equation}
While it will turn out that in the equilibrium case the first moment of $f$ 
in Eq. (\ref{31}) may be identified with the quantity $N ^{i} = n u ^{i}$ introduced below Eq. (\ref{6}), we have used here the same symbol  immediately. 
The second moment of $f$ denoted by $\tilde{T}^{ik}$ in Eq. (\ref{31}) 
will not, however, coincide with the energy-momentum tensor $T ^{ik}$ in 
Eq. (\ref{8}). 
The integrals in the definitions (\ref{31}) and in the following  
are integrals over the entire mass shell 
$p^{i}p_{i} = - m^{2}$. 
The entropy flow vector $S^{a}$ is given by \cite{Ehl}, \cite{IS} 
\begin{equation}
S^{a} = - \int p^{a}\left[
f\ln f - f\right]\mbox{d}P \mbox{ , }
\label{32}
\end{equation}
where we have restricted ourselves to the case of 
classical Maxwell-Boltzmann particles. 
 
Using well-known general relations (see, e.g., \cite{Stew})  
we find 
\begin{eqnarray}
N^{a}_{;a} &=& \int \left(C\left[f\right] 
- m F ^{i}\frac{\partial{f}}{\partial{p ^{i}}}\right) \mbox{d}P 
\mbox{ , } \ \ \nonumber\\
\tilde{T}^{ak}_{\ ;k} &=&  \int p^{a}\left(C\left[f\right] 
- m F ^{i}\frac{\partial{f}}{\partial{p ^{i}}}\right) 
\mbox{d}P
\mbox{ , } 
\label{33}
\end{eqnarray}
and 
\begin{equation}
S^{a}_{;a} = - \int \ln f 
\left(C\left[f\right] 
- m F ^{i}\frac{\partial{f}}{\partial{p ^{i}}}\right) \mbox{d}P
\mbox{ .} 
\label{34}
\end{equation}
In the following we will focus on the force terms in the expressions (\ref{33}) and (\ref{34}). 
In order to separate the corresponding contributions from those due to the collision integral we will restrict ourselves to collisional equilibrium from now on. 
Under this condition  $\ln f$ in Eq. 
(\ref{34}) 
is a linear combination of the collision invariants 
$1$ and $p^{a}$ and the contributions due to $C \left[f \right]$ in formulas (\ref{33}) and (\ref{34}) vanish.  
The corresponding equilibrium distribution function 
becomes (see, e.g., \cite{Ehl}) 
\begin{equation}
f^{0}\left(x, p\right) = 
\exp{\left[\alpha + \beta_{a}p^{a}\right] } 
\mbox{ , }\label{35}
\end{equation}
where $\alpha = \alpha\left(x\right)$ and 
$\beta_{a}\left(x \right)$ is timelike.\\ 
Inserting the equilibrium distribution function (\ref{35}) 
into Eq.(\ref{29}) one obtains
\begin{equation}
p^{a}\alpha_{,a} +
\beta_{\left(a;b\right)}p^{a}p^{b}   
=  - m \beta _{i}F ^{i} 
\mbox{ .} \label{36}
\end{equation} 
For a vanishing force $F ^{i}$ the latter condition reduces to the  ``global'' equilibrium condition of standard relativistic kinetic theory. 
For $F ^{i} \neq 0$ condition (\ref{36}) is a ``generalized'' equilibrium condition (see \cite{Zsub} and below). 

Use of Eq. (\ref{35}) in the balances (\ref{33}) yields 
\begin{eqnarray}
N^{a}_{;a}&=&-m \beta _{i}\int F ^{i}f ^{0}\mbox{d}P \ , \nonumber\\ 
\tilde{T}^{ak}_{\ ;k}&=&-m \beta _{i}\int p^{a}F ^{i}f ^{0}\mbox{d}P \ .
\label{37}
\end{eqnarray}

For the entropy production density (\ref{34}) we find 
\begin{eqnarray}
S^{a}_{;a} &=& m \beta _{i} \int \left[\alpha + \beta _{a}p ^{a} \right]
F ^{i}
\ln f^{0} \mbox{d}P \nonumber\\
&=&  - \alpha N^{a}_{;a} 
- \beta_{a}\tilde{T}^{ab}_{\ ;b}
\mbox{ . }
\label{38}
\end{eqnarray}  

With $f$ replaced by $f^{0}$ in the definitions 
(\ref{31}) and (\ref{32}), $N^{a}$, $\tilde{T}^{ab}$ and $S^{a}$ may be 
split with respect to the unique 4-velocity $u^{a}$ according to 
\begin{equation}
N^{a} = nu^{a} \mbox{ , \ \ }
\tilde{T}^{ab} = \rho u^{a}u^{b} + p h^{ab} \mbox{ , \ \ }
S^{a} = nsu^{a} \mbox{  . }
\label{39}
\end{equation}
where $u ^{a}$, $h ^{ab}$,  
$n$,  $\rho$, $p$ and 
$s$ may be identified with the corresponding quantities of the previous section.  
The exact integral expressions for $n$, $\rho$, $p$ and $s$ are given
by the formulae (177) - (180) in \cite{Ehl}. 

Using the structure (\ref{39}) for $N ^{a}$ and defining
\begin{equation}
\Gamma  \equiv -\frac{m}{n} \beta _{i}\int F ^{i}f ^{0} \mbox{d}P 
\mbox{ , }
\label{40}
\end{equation}
the first Eq.(\ref{37}) becomes 
$\dot{n} + \Theta n = n\Gamma$ [cf. Eq. (\ref{7})] which justifies definition (\ref{40}).  
Similarly, 
with the decomposition (\ref{39}) and the abbreviation
\begin{equation}
t^{a} \equiv m \beta _{i} \int p^{a}F ^{i}f ^{0} \mbox{d}P \mbox{ , }
\label{41}
\end{equation}
we obtain 
\begin{equation}
\tilde{T} ^{ab}_{\ ;b} + t ^{a} = 0 \ ,
\label{42}
\end{equation}
implying
\begin{eqnarray}
\dot{\rho } + \Theta \left(\rho + p \right) &=& u _{a}t ^{a}\ ,
\nonumber\\
\left(\rho + p \right)\dot{u}_{a} + \nabla  _{a}p 
&=& - h _{ai}t ^{i}\ .
\label{43}
\end{eqnarray}
The energy-momentum tensor $\tilde{T}^{ik}$ is not conserved. 
Obviously, the balances (\ref{43}) following from Eq. 
(\ref{42}) are identical to the balances 
\begin{eqnarray}
\dot{\rho } + \Theta \left(\rho + p + \pi  \right) &=& 0 \ ,
\nonumber\\ 
\left(\rho + p + \pi  \right)\dot{u}_{a} 
+ \nabla  _{a}\left(p + \pi  \right) &=& 0
\label{44}
\end{eqnarray}
following from the local conservation $T^{ab}_{\ ;b} = 0$ of an effective energy-momentum tensor $T ^{ab}$, 
\begin{equation}
T^{ab} = \rho u ^{a}u ^{b} + \left(p + \pi  \right)h ^{ab}
\label{45}
\end{equation}
with the identifications
\begin{equation}
u _{a}t ^{a} = - \Theta \pi \ ,
\ \ \ \ \ 
h _{ai}t ^{i} = \pi \dot{u}_{a} + \nabla  _{a}\pi \ .
\label{46}
\end{equation}
This mapping of the ``source'' term $t ^{a}$ onto an effective viscous pressure $\pi $ of a locally conserved energy-momentum tensor $T ^{ab}$ was 
explicitly shown to be consistent for specific ``sources'' $t ^{a}$, depending on the first and second moments of the distribution function only \cite{TZP,ZTP,Zsub}. 
In the following subsection we will show that this interpretation continues to hold if specific third moments are involved. 

Let us now decompose the four-momenta $p ^{a}$ into 
$p ^{a} = E u ^{a} + \lambda e ^{a}$ where $e ^{a}$ is a unit spatial vector, i.e., $e ^{a}e _{a} = 1$, 
$e ^{a}u _{a} = 0$. 
Consequently, one has $E = - u _{a}p ^{a}$ and $\lambda = e _{a}p ^{a}$ 
and the 
mass shell condition $p ^{a}p _{a} = - m ^{2}$ is equivalent to 
$\lambda ^{2} = E ^{2} - m ^{2}$. 
Moreover, $h _{ab}p ^{a}p ^{b} = \lambda ^{2}$ is valid.
For the force $F ^{m}$ we write analogously
$F ^{m} = F \left(x,p \right)u ^{m} 
+ K \left(x,p \right)e ^{m}$
where $F \equiv  - u _{m}F ^{m}$ and $K \equiv  e _{m}F ^{m}$. 
The requirement $p _{m}F ^{m} = 0$ will be automatically fulfilled for 
\begin{equation}
F ^{m} = \left[u ^{m} + \frac{E}{\lambda }e ^{m} \right]
F \left(x,p \right)\ .
\label{47}
\end{equation}
With the familiar identification $\beta _{m} = u _{m}/T$ it is obvious that only the part $u _{m}F ^{m} \equiv  - F$ contributes in the ``sources'' 
(\ref{40}) and 
(\ref{41}). 

\subsection{Quadratic self-interaction}
Any specific force  relies on reasonable assumptions about 
$F \left(x,p \right)$. 
In a previous paper \cite{ZiBa} we investigated the most general linear dependence of $F$ on the particle momentum $p ^{a}$. 
In this paper we assume $F$ to depend quadratically on the spatially projected four-momenta $\lambda $, i.e., 
\begin{equation}
F \left(x,p \right) \equiv  - u _{i}F ^{i} 
= \lambda F _{1}\left(x \right) + \lambda ^{2}F _{2}\left(x \right)\ .
\label{48}
\end{equation}
$F _{1}\left(x \right)$ and $F _{2}\left(x \right)$ are spacetime functions to be determined by the equilibrium conditions of the gas. 
It will turn out that it is just this force under the action of which the particle energy $E$ is preserved in a homogeneous and isotropic universe 
[cf. Eq. (\ref{71}) below]. \\
An equivalent way of writing the force (\ref{47}) with $F$ given by Eq. (\ref{48}) 
is 
\begin{equation}
F ^{i}\left(x,p \right) = g _{ab}p ^{a}
\left[u ^{i}e ^{b} - u ^{b}e ^{i} \right]
\left[F _{1}\left(x \right) + \lambda F _{2}\left(x \right)\right]\ .
\label{49}
\end{equation}
The equilibrium condition (\ref{36})  becomes 
\begin{equation}
p ^{a}\alpha _{,a} + \beta _{\left(a;b \right)}p ^{a}p ^{b} = 
\frac{m}{T}
\left[e _{a}p ^{a} F _{1}\left(x \right) + h _{ab}p ^{a}p ^{b}
F _{2}\left(x \right)  \right]\ .
\label{50}
\end{equation}
It is satisfied for 
\begin{equation}
\dot{\alpha}= 0\ ,\ \ \ \ 
e ^{a}\nabla  _{a}\alpha = \frac{m}{T}F _{1}\ ,
\label{51}
\end{equation}
and 
\begin{equation}
\beta_{(a;b)}= \phi (x)h_{ab}\ , 
\label{52}
\end{equation}
with $\phi  = \frac{m}{T} F _{2}$. 
Since, on the other hand, relation (\ref{2c}) holds we find 
\begin{equation}
F _{2}\left(x \right) = \frac{\Theta }{3m}\ .
\label{53}
\end{equation}

Identifying $\beta _{a}$ with $u _{a}/T$, 
{\it the equilibrium condition 
(\ref{52}) for a Maxwell-Boltzmann gas in the force field (\ref{49}) 
coincides with the PCTKV condition (\ref{5})}. 

We recall that an ansatz for $F$ linear in the particle momenta instead of the structure (\ref{48}) analogously reproduces the conformal Killing-vector condition (\ref{2g}) \cite{ZiBa}. 

Through $F _{2}$ the force depends on the fluid expansion, a quantity characterizing the gas as a whole on the macroscopic level. 
Since both the microscopic particle momenta and macroscopic fluid quantities of the system of gas particles enter the four-force, the latter 
represents a self-interaction of the gas. 

Inserting the force (\ref{49}), equivalent to the expressions (\ref{47}) and (\ref{48}), into the source term (\ref{40}) and using 
$3 p = h _{ab}\tilde{T}^{ab} = 3nT$ we obtain 
\begin{equation}
\Gamma = 3m F _{2}\left(x \right)\ .
\label{54}
\end{equation}
Together with $F _{2}$ from Eq. (\ref{53}) this implies $\Gamma = \Theta $ [cf. Eq. (\ref{17})] and, consequently, the relation $\dot{n}/n = 0$.  

For the source term  (\ref{41}) one obtains
\begin{equation}
t ^{a} = - \frac{m}{T}F _{1}\left(x \right)e _{i}\tilde{T}^{ai} 
- \frac{m}{T}F _{2}\left(x \right)h _{im}M ^{aim}\ ,
\label{55}
\end{equation}
where $M ^{aim} \equiv   \int \mbox{d}Pf ^{0}p ^{a}p ^{i}p ^{m}$ is the third moment of the equilibrium distribution function. 
The projected source terms in the balances (\ref{43}) are 
\begin{equation}
u _{a}t ^{a} = 3 \eta \frac{m}{T}F _{2}\left(x \right)\ ,
\label{56}
\end{equation}
where we have introduced the quantity
\begin{equation}
\eta \equiv  - \frac{1}{3}u _{a}h _{im}M ^{aim}\ ,
\label{57}
\end{equation}
and 
\begin{equation}
h _{na}t ^{a} = - e _{n}p \frac{m}{T} F _{1}\left(x \right)\ .
\label{58}
\end{equation}
The third moment $M ^{aim}$ enters the energy balance but not the momentum balance. 
The momentum balance in Eq. (\ref{43}) with the ``source'' term (\ref{58}) coincides with the corresponding balance in \cite{Zsub}. 
It follows that the arguments in \cite{Zsub} which prove the consistency of the momentum balances in Eqs. (\ref{43}) and (\ref{44}) together with the second relation in Eq. (\ref{46}) apply in the present case as well. 
Realizing that 
\begin{equation}
u _{a}p ^{a}f ^{0} = \frac{\partial{f ^{0}}}
{\partial \left({\frac{1}{T}} \right)}
\label{59}
\end{equation}
(the derivative has to be taken for $\alpha = const$), we find 
\begin{equation}
\eta = - \frac{1}{3}h _{mn}\frac{\partial \tilde{T}^{mn}}
{\partial \left(\frac{1}{T} \right)} 
= - \frac{\partial{p}}{\partial \left(\frac{1}{T} \right)}\ .
\label{60}
\end{equation}
Using here for the equilibrium pressure $p$ \cite{Groot}
\begin{equation}
p = \frac{4 \pi m ^{4}}{\left(2 \pi  \right)^{3}}
\frac{K _{2}\left(\frac{m}{T} \right)}{\left(\frac{m}{T} \right)^{2}}
\exp{\left[\alpha \right]}
\label{61}
\end{equation}
together with 
\begin{equation}
\frac{\mbox{d}}{\mbox{d}z}\left(\frac{K _{2}\left(z \right)}{z ^{2}} \right) 
= - \frac{K _{3}\left(z \right)}{z ^{2}}
\label{62}
\end{equation}
($K _{2}$ and $K _{3}$ are Bessel functions of the second kind), one obtains 
\begin{equation}
\eta = nTm \frac{K _{3}\left(\frac{m}{T} \right)}
{K _{2}\left(\frac{m}{T} \right)}
= T \left(\rho + p \right)\ ,
\label{63}
\end{equation}
where we took into account that the enthalpy per particle $h$ is given by 
$h = \left(\rho + p \right)/n = 
m K _{3}\left(\frac{m}{T} \right)/K _{2}\left(\frac{m}{T} \right)$. 
Consequently, the source term $u _{a}t ^{a}$  becomes
\begin{equation}
u _{a}t ^{a} = 3m \left(\rho + p \right)F _{2}\left(x \right) 
= \left(\rho + p \right)\Theta \ .
\label{64}
\end{equation}
Together with the first relation in Eq. (\ref{46}) this is consistent with 
$\pi = - \left(\rho + p \right)$ [cf. Eq. (\ref{17})]. 

With the identifications $\alpha = \mu /T$, $\beta _{a} = u _{a}/T$ and  
$s = \frac{\rho + p}{nT} - \frac{\mu }{T}$, the entropy production density (\ref{38}) may be expressed in terms of the sources $\Gamma $ and $u _{a} t ^{a}$ to yield 
\begin{equation}
S ^{a}_{;a} - ns \Gamma = - \frac{\rho + p}{T}\Gamma 
+ \frac{u _{a}t ^{a}}{T} = 0 \ .
\label{65}
\end{equation}

Recalling that Eq. (\ref{52}) implies $\Gamma = \Theta$ and $\dot{n} = 0$, 
we have {\it derived} all the generalized equilibrium conditions discussed on a phenomenological level in the previous section. 
It is remarkable that the force (\ref{47}) with the quadratic dependence 
(\ref{48}) does not only lead to conditions 
(\ref{52}) but automatically guarantees the vanishing of 
$S ^{a}_{;a} - ns \Gamma $, i.e. $\dot{s} = 0$, resulting in 
$\dot{T}/T = 0$ according to the temperature law (\ref{15}). 
{\it The condition $\dot{s} = 0$ for ``adiabatic'' or ``isentropic'' particle production 
is a property of the force considered here and needs not to be postulated separately.}

For any $\Gamma = \Theta \geq 0$ Eq. (\ref{65}) implies  
$S ^{a}_{;a} \geq 0$. {\it The self-interacting force provides us with a nonnegative expression for the entropy production in an expanding universe.} 

This completes our microscopic derivation of generalized equilibrium 
characterized by Eqs. 
(\ref{13}) and (\ref{16}) under the PCTKV condition (\ref{5}). 

\section{The self-interacting gas universe}
Having clarified the implications of generalized equilibrium, especially the consequences of relation (\ref{5}), both phenomenologically and on the level of kinetic theory, we now consider Einstein's field equations with the energy-momentum tensor (\ref{45}). 
This corresponds to a situation where matter in generalized equilibrium dominates the dynamics of the universe. 
Since generalized equilibrium was shown to be realized by a self-interacting gas we call such kind of configuration a self-interacting gas universe. 
While it is generally an open question to what extent the hot and dense early universe is accessible to a kinetic description, a gas is the only system for which the correspondence between microscopic variables and phenomenological fluid quantities is sufficiently well understood. 
This makes gas universes interesting toy models and we hope that such kind of approach also gives an idea of the relevant physics in our real universe. 

The target of this section is to demonstrate explicitly, that the specific self-interaction (\ref{49}) (equivalent to the combination of Eqs. (\ref{47}) and (\ref{48})) under the equilibrium conditions (\ref{51}) and (\ref{52}) allows us to exactly integrate both the equations for the cosmic scale factor of a homogeneous and isotropic self-interacting gas universe and the corresponding microscopic equations of motion for the individual gas particles. 
A self-interacting gas universe represents an exactly solvable model both microscopically and on the phenomenological fluid level. 

In the spatially homogeneous case the function $F _{1}$ vanishes 
[cf. Eq. (\ref{51})]. 
The lenght scale $a$ coincides with the scale factor of the Robertson-Walker metric and obeys the equations 
\begin{equation}
3 \frac{\dot{a}^{2}}{a ^{2}} = \kappa \rho \ ,
\label{65a}
\end{equation}
where $\kappa$ is Einstein's gravitational constant, 
and 
\begin{equation}
\left(\frac{\dot{a}}{a} \right)^{\displaystyle \cdot} 
= - \frac{\kappa}{2}\left(\rho + p + \pi  \right)
\ .
\label{65b}
\end{equation}
Together with the expression (\ref{17}) for $\pi $ the last equation yields 
$\dot{a}/a \equiv  H = const$, where $H$ is the Hubble parameter,  
implying an exponential behaviour of the scale factor, 
$a \propto \exp{\left[Ht \right]}$.  

{\it A homogeneous and isotropic simple gas universe with arbitrary equation of state and 
quadratic (in the spatially projected microscopic particle four-momenta) self-interaction inbetween elastic binary collisions requires a de Sitter spacetime to be in (generalized) equilibrium.}

Evidently, this also implies that the de Sitter metric admits 
a PCTKV.  The consistency of this statement may be checked from Eq. (\ref{5}) directly by using 
that the temperature $T$ is constant both in space and time, together with 
the well-known decomposition of the covariant derivative of the 
four-velocity \cite{Ehlers,Ellis}, 
\begin{equation}
u _{i;n} = - \dot{u}_{i}u _{n} + \sigma _{in} +  \omega _{in}  
+ \frac{\Theta }{3}h _{in}\ ,
\label{65c}
\end{equation}
where
$\omega_{ab} = h_{a}^{c}h_{b}^{d}u_{\left[c;d\right]}$. 
The present case is characterized by 
$\dot{u}_{a} = \sigma _{ab} = \omega _{ab} = 0$ and 
$\Theta = 3 \frac{\dot{a}}{a} = const$ as well as 
$\phi = \frac{\Theta }{3 T} = const$ [cf. Eq. (\ref{2c})].  

In other words, 
{\it the selfinteraction of a classical gas is able to realize an effective fluid equation of state $P _{eff} \equiv  p + \pi = - \rho $} 
[cf. Eq. (\ref{17})] which in a cosmological context is usually obtained with the help of a scalar field. 
A scalar field represents an ``exotic'' kind of matter with mainly theoretical evidence at the present state of knowledge. 
We argue here that it may occasionally be helpful  to have alternative ways of considering issues of inflation in terms of conventional matter models which are more familiar and intuitive compared with the ``exotic'' ones. 
A remarkable difference to scalar-field-driven inflation is the circumstance that our approach predicts an exponential increase of the comoving entropy 
$nsa ^{3}$ during the de Sitter phase. 

Our results may also shed new light on the old question whether or not a fluid bulk pressure may drive inflation. 
We recall that this issue has been discussed in the literature from different points of view (\cite{Pach,LiPoWa,PaBaJ,HiSa,ZaJ,Roy,WZ,RM}). 
While Pacher et al.\cite{Pach} have shown that sufficiently high negative pressures cannot arise in a weakly interacting mixture of relativistic and nonrelativistic particles (see also \cite{HiSa}), Lima et al. \cite{LiPoWa} pointed out that the situation may be different if the dilute-gas approximation is given up and causal thermodynamics is applied. 
Further investigations along this line have confirmed the existence of inflationary solutions \cite{PaBaJ,ZaJ,Roy,WZ,RM}, although there exist general problems with their physical interpretation \cite{Roy,WZ,RM}, at least as long as cosmological particle production is not taken into account. 
The fact that a bulk pressure may phenomenologically represent certain quantum 
phenomena, especially particle production processes, is well known in the literature \cite{Zel,Mur,Hu}. 
As was remarked in \cite{LiGer,GaLeDe} ``conventional'' bulk pressures, i.e. bulk pressures due to internal interactions, and effective bulk pressures resulting from particle production are separate effects and both of them contribute to the overall dynamics of the system. 

We emphasize again that 
the quantity $\pi $ in the present paper is exclusively due an increase in the number of particles and {\it not} the ``conventional'' bulk viscous pressure of linear, irreversible thermodynamics, describing internal particle number preserving interactions. 
``Conventional'' bulk pressures, generally equivalent to 
$\dot{s}\neq 0$, have been excluded here by the assumption of collisional equilibrium. 
The possibility of a nonvanishing $\dot{s}$ due to the process of particle production was eliminated by the requirement of ``generalized'' equilibrium 
(see the discussion below Eq. (\ref{13})). 

The above mentioned studies within the framework of causal thermodynamics relied on deviations from thermodynamical equilibrium characterized by 
$\dot{s} \neq 0$ whereas the present considerations refer to a (generalized) equilibrium and imply $\dot{s} = 0$. 
While there are limits for deviations from equilibrium in the mentioned nonequilibrium appoaches (deviations up to second order), there are no such  restrictions in the context of this paper. 
In particular, there is no need of a requirement $| \pi | < p$ in our case to be well within the range of applicability of the theory as in conventional nonequilibrium thermodynamics. 
The present analogue of the bulk viscous pressure $\pi $ which in conventional irreversible thermodynamics represents (small) deviations from (collisional) equilibrium is a quantity without corresponding limitations. Instead 
it is determined by equilibrium conditions, equivalent to symmetry requirements. 
This quantity $\pi $ is directly related to the particle production rate   
which is traced back to a simple force on the (classical) microscopic level. 
It follows that $\pi $ is completely determined by this force. 
The problem whether a bulk pressure may drive inflation reduces to the question whether there exist microscopic forces on the particles, equivalent to a nongeodesic motion of the latter, which generate an appropriate macroscopic quantity $\pi $. 
Our considerations show that a surprisingly simple force generates such a quantity. 
In this sense the question whether or not an effective bulk pressure may drive inflation is answered affirmatively. 
A final statement, however, requires the derivation of this force from an underlying quantum level which is beyond the scope of this paper. 

Having determined the selfinteracting force by the equilibrium conditions of the gas it is now also possible to study the particle motion (\ref{30}) explicitly. 
With the decomposition $p ^{i} = E u ^{i} + \lambda e ^{i}$ the left-hand side of the second equation (\ref{30}) may be written as 
\[
\frac{D p ^{i}}{d \tau } = \frac{d E}{d \tau }u ^{i} 
+ E \frac{D u ^{i}}{d \tau } + \frac{d \lambda }{d \tau }e ^{i} 
+ \lambda \frac{D e ^{i}}{d \tau }\ .
\]
Contraction with $u _{i}$ yields 
\[
u _{i}\frac{D p ^{i}}{d \tau } = - \frac{d E}{d \tau }   
+ \lambda u _{i}\frac{D e ^{i}}{d \tau } 
= - \frac{d E}{d \tau }   
- \lambda e ^{i}\frac{D u _{i}}{d \tau }\ .
\]
Taking into account that 
\[
\frac{D u ^{i}}{d \tau } = u ^{i}_{;n}\frac{p ^{n}}{m}\ ,
\]
we obtain 
\[
u _{i}\frac{D p ^{i}}{d \tau } = - \frac{d E}{d \tau }   
- \frac{\lambda E}{m}e ^{i}\dot{u}_{i} 
- \frac{\lambda ^{2}}{m}e ^{i}e ^{n}u _{i;n}\ .
\]
Applying here the decomposition (\ref{65c}),  
the projected equation of motion 
\begin{equation}
u _{i}\frac{D p ^{i}}{d \tau } = u _{i}F ^{i} = - F
\label{66}
\end{equation}
may generally be written as 
\begin{equation}
\frac{d E}{d \tau } + \frac{\lambda E}{m}e ^{i}\dot{u}_{i} 
+ \frac{\lambda ^{2}}{m}e ^{i}e ^{n}\sigma _{in} 
+ \frac{\lambda ^{2}}{3m}\Theta  = F \ .
\label{67}
\end{equation}
For homogeneous, isotropic universes with $\dot{u}_{i} = \sigma _{in} = 0$ the last equation reduces to 
\begin{equation}
\frac{d E}{d \tau } + \frac{\lambda ^{2}}{3m}\Theta  = F \ .
\label{68}
\end{equation}
With $d \tau = d t \left(m/E \right)$, $\lambda ^{2} = E ^{2} - m ^{2}$, 
$\Theta = 3 \dot{a}/a$ and $d E/d t \equiv  \dot{E}$, Eq. (\ref{68}) is equivalent to 
\begin{equation}
\frac{\left(E ^{2} - m ^{2} \right)^{\displaystyle \cdot}}
{E ^{2} - m ^{2}} + \frac{\left(a ^{2} \right)^{\displaystyle \cdot}}
{a ^{2}} = \frac{2m}{E ^{2} - m ^{2}}F \ .
\label{69}
\end{equation}
We discuss the last equation for three different cases: 

(i) $F = 0$, geodesic motion. 
We find   
\begin{equation}
E ^{2} - m ^{2} = \lambda ^{2} \propto a ^{-2}\ ,
\mbox{\ \ \ }
\left(F = 0 \right)\ ,
\label{70}
\end{equation}
implying the expected behavior $E \propto a ^{-1}$ for massless 
particles (photons) while the nonrelativistic energy $\epsilon \equiv  E - m$ with $\epsilon \ll m$ of massive particles decays as $\epsilon \propto a ^{-2}$. 

(ii) $F = \left(E - m \right)\Theta /3$. 
The previously studied force with a linear dependence of $F$ on the particle four-momenta is given by this expression for massive particles ($m \gg T$) in a homogeneous universe \cite{ZiBa}. 
In such a case, which implies the conformal Killing-vector property (\ref{2g}) 
and power-law inflation according to $a \propto t ^{4/3}$ \cite{ZiBa}, the solution of Eq. (\ref{68}) is 
\begin{equation}
E - m  \propto  a ^{-1} \ ,
\mbox{\ \ \ \ \ }
\left(F = \left(E - m \right)\frac{\Theta }{3} \right)\ . 
\label{70a}
\end{equation}
The nonrelativistic energy of massive particles under generalized equilibrium conditions in a (quasi-)linear force field decays linearly with the cosmic scale factor, i.e., in this case the selfinteracting force makes nonrelativistic particles behave like radiation. 
This may be regarded as the microscopic counterpart of the statement that radiation and nonrelativistic matter may be in equilibrium in the expanding universe, provided the number of matter particles increases at a specific rate 
\cite{ZTP,ZiBa}.

(iii) $F = \frac{\lambda ^{2}}{3m} \Theta$. 
This is the case of interest here [cf. Eq. (\ref{48}) with 
$F _{1} = 0$ and $F _{2}$ from Eq. (\ref{53})]. 
It is obvious from Eq. (\ref{68}) that the force term on the right-hand side 
exactly compensates the second term on the left-hand side. 
Consequently, $\dot{E}$ vanishes, i.e. 
\begin{equation}
E = const \ ,
\mbox{\ \ \ \ \ }
\left(F = \frac{\lambda ^{2}}{3m} \Theta \right)\ .
\label{71}
\end{equation} 
The self-interacting force prevents the particle energies from decaying with the expansion. 
Independently of the equations of state the particle energies are preserved in such a universe. 

With the result (\ref{71}) we have completed the exact solution of our model of a quadratically self-interacting gas universe. 
It is the essential feature of this model that 
{\it the same force which on the microscopic level makes the gas particles move at constant energy is responsable for an effective gravitational repulsion on the macroscopic level, implying an exponentially accelerated expansion of the universe.}

\section{Conclusions}
In this paper we introduced the concept of a ``projector-conformal'' timelike  Killing vector (PCTKV) and discussed the corresponding fluid dynamics under the condition of minimal entropy production (generalized equilibrium). 
Such kind of equilibrium configuration requires a particle production rate which coincides with the fluid expansion rate. 
As a consequence the energy density of the fluid turned out to be stationary. 
A microscopic realization of this phenomenologically defined concept was given with the help of the kinetic theory for a classical gas in a force field. 
A quadratic (in the particle four-momenta) self-interaction of the microscopic gas particles was shown to provide both the PCTKV property of $u _{i}/T$ and 
``adiabatic'' (or ``isentropic'') particle production. 
This force concept turned out to result in a comprehensive picture of the gas dynamics both macroscopically and microscopically and allowed us to establish an exactly solvable model of a quadratically self-interacting gas universe. 
We found that generalized equilibrium under the conditions of spatial homogeneity and isotropy for such a configuration necessarily implies a de Sitter spacetime. 
We clarified in which sense an effective bulk pressure may drive exponential 
inflation.

\acknowledgments
This paper was supported by the Deutsche Forschungsgemeinschaft.

\end{document}